# Increasing Herd Immunity with Influenza Revaccination

Eric Mooring[1,4], Shweta Bansal [1,2,3]

1. Department of Biology, Georgetown University, Washington DC 20057, USA

2. Fogarty International Center, National Institutes of Health, Bethesda MD 20892, USA

3. shweta@sbansal.com

4. Current address: Center for Communicable Disease Dynamics, Department of Epidemiology, Harvard T.H. Chan School of Public Health, Boston MA 02115, USA

**Abstract**

Seasonal influenza is a significant public health concern in the United States and globally. While influenza vaccines are the single most effective intervention to reduce influenza morbidity and mortality, there is considerable debate surrounding the merits and consequences of repeated seasonal vaccination. Here, we describe a two-season influenza epidemic contact network model and use it to demonstrate that increasing the level of continuity in vaccination across seasons reduces the burden on public health. We show that revaccination reduces the influenza attack rate not only because it reduces the overall number of susceptible individuals, but also because it better protects highly-connected individuals, who would otherwise make a disproportionately large contribution to influenza transmission. Our work thus contributes a population-level perspective to debates about the merits of repeated influenza vaccination and advocates for public health policy to incorporate individual vaccine histories.

**Introduction**

Influenza is a serious public health threat in the United States. In recent decades, seasonal influenza has contributed to about 30,000 excess deaths per year on average [1], is a significant cause of outpatient and emergency department visits by children [2-3], and accounts for millions of lost workdays each year [4]. Controlling influenza is a multifaceted effort utilizing strategies such as antiviral drug treatment and prophylaxis [4] and hygiene practices like washing hands and wearing face masks [5-6]. However, seasonal influenza vaccination has been the centerpiece of influenza control efforts in the United States for the past 60 years and vaccine coverage of only 40% is believed to reduce the risk of influenza illness by about 60% among the overall population [7].

While the impact of overall influenza vaccine coverage levels in a single season has been studied in detail [8-12], the consequences of repeated seasonal influenza vaccination to the same individuals have not been studied as extensively. Most research on repeated vaccination has considered this problem from an immunological [13-14] or purely statistical perspective and not considered potential population-level consequences of repeated vaccination. For example, Ohmit et al. [15] recently found that seasonal influenza vaccines are ineffective for patients who were vaccinated the previous season, but Keitel et al. [16] and Voordouw et al. [17] found that repeated vaccination contributes to increased protection. Such discrepancies may be explained by the antigenic distance hypothesis, which depends on positive and negative interference between vaccine strains [18].

A thorough understanding of the merits of vaccination strategies may require considering their consequences across multiple seasons. Carrat et al. [19] assume no long-term immunological benefits from repeated seasonal influenza vaccination, and model long-run effects of vaccination strategies which prescribe vaccinating adults and children, such as the strategy recently adopted in the United States. Their analysis suggests that vaccinating people throughout their lives prevents them from developing natural immunity to influenza and therefore increases the risk of infection at older ages. Counterintuitively, their results suggest that given the greater risk of mortality associated with influenza at increasing ages, vaccination at all ages may actually increase influenza mortality. However, their analysis does not directly model influenza epidemics and does not account for any herd immunity impacts of higher vaccination rates or rates of individual-level revaccination. Similarly, Bodewes et al. [20] suggest that repeated seasonal influenza vaccination may be disadvantageous because it might prevent patients from developing heterosubtypic immunity that could potentially be protective during a pandemic [21].

Mathematical epidemiological models have been used extensively to study vaccination and other influenza control strategies, and consider targeted vaccination to minimize morbidity, mortality, or economic costs [22-27]. Some mathematical models as well as epidemiological data suggest that targeting influenza vaccination towards school-age children may be a preferred strategy, as children are the age group most likely to be infected with influenza and to transmit it to others [22, 26]. Others have advocated for a strategy that minimizes mortality by targeting those most at risk for complications and death [24, 27]. However, most models of influenza vaccination have focused on single epidemics and hence not accounted for the rate of revaccination (but see Fung et al. [28]). Also, modeling revaccination inherently requires a modeling framework such as contact network modeling that explicitly models individual hosts. Recent theoretical research using contact network models has shown the significance of modeling epidemics in series, when natural immunity from past epidemics influences future ones [29-30].

In this study, we consider whether the rate of revaccination, which we define as the proportion of first season vaccine recipients who are vaccinated in a successive influenza season, may be epidemiologically relevant. We present a two-season mathematical model to explore the consequences of influenza revaccination on herd immunity and the mechanisms driving them. Our theoretical study suggests that revaccination indeed reduces the public health burden, and that this result is robust with respect to variation in contact structure, vaccine efficacy, vaccine coverage, vaccine assortativity, and levels of natural immunity. Our work thus contributes a population-level perspective to debates about the merits of repeated influenza vaccination and advocates for public health policy to incorporate individual vaccine histories.

**Results**

We define the rate of *revaccination* ($r$) as the proportion of vaccine recipients in one influenza season who are also vaccinated in the following season. If vaccination were random with respect to previous vaccination status, the expected revaccination rate would be equal to the vaccination coverage. Because the expected revaccination rate is a function of the vaccination coverage, we additionally define $r'$ to be the *excess revaccination rate*, which takes into account variable vaccination coverage and measures revaccination beyond the level expected at random. (Details in Methods.) However, for almost all of our analyses we will simply use $r$, because vaccination coverage will be held constant.

To assess the epidemiological relevance of revaccination rates, we present a contact network model for two consecutive influenza seasons. In the contact network model, each individual is represented as a node, and influenza-spreading contacts or interactions are represented as edges (Figure 1). Prior to the first influenza season, all individuals in the population are susceptible to infection. A proportion of this population is protected by pre-season vaccination. We assume that vaccine-induced immunity is fully-protective in a season and is protective only in the season in which vaccination occurs, but that natural immunity confers protection in the season following infection (further information in Methods). Outbreaks are simulated until a large epidemic (i.e. ≥5% of individuals infected) occurs. Following a first season epidemic, vaccination is implemented again prior to the second season outbreak. The identity of the second season vaccine recipients is chosen based on the revaccination rate ($r$, ranging from 0% to 100%), and is implemented randomly. We assume that the level of vaccination coverage is constant across both seasons, supported by the National Health Interview Survey which shows that flu vaccination coverage in the United States has been quite consistent from the 2007-2008 season to the 2011-2012 season [31]. We record results from second seasons in which a large epidemic occurs. We focus on the size of second-season epidemics, because the rate of revaccination inherently cannot have any bearing on first season outbreaks.

When the model is applied to a synthetic (computationally-generated) exponential random contact network (details in Methods below), we find that second season epidemic sizes decrease as the rate of revaccination increases (Figure 2). This result indicates that the rate of revaccination is indeed epidemiologically relevant, and we explore below some of the mechanisms leading to this effect.

*Explaining the Effect of Revaccination: "Wasting" Vaccine*

One possible cause of the effect of revaccination on epidemic size is that as the rate of revaccination increases, vaccine is used more efficiently because individuals with natural immunity from first-season infection are less likely to also be vaccinated for the second season. When revaccination is complete ($r = 100\%$), no individual can be both naturally immune (i.e. protected due to first season infection) and have second-season vaccination because, under total revaccination, all individuals vaccinated immediately prior to the second season must have also been vaccinated and thus fully protected for the first season. Under the assumptions of the model, vaccinating individuals with natural immunity does not provide further protection to those individuals or in terms of herd immunity to the population. As a result, more revaccination leads to fewer susceptible individuals prior to the second season, which indicates a more efficient use of vaccine.

We test this by comparing the results of two models that differ only in terms of which individuals are eligible to be vaccinated for the first time prior to the second season. One model—universal vaccination—follows the method outlined in the previous section, in which vaccine doses not used for revaccination are randomly given to any individuals not vaccinated for the first season, including both naturally immune and never vaccinated or infected individuals. An alternative model—preferential vaccination—still implements revaccination as before, but the only individuals who are candidates for first-time vaccination prior to the second season are those without natural immunity from the first season, thus not "wasting" vaccine on any naturally-immune individuals. While this approach may not be realistic because individuals who are infected with influenza in a previous season might be more motivated to be vaccinated in a subsequent season, it provides a useful model for comparison. When revaccination is complete ($r = 100\%$), no individuals are vaccinated for the first time prior to the second season and, as such, the two models are functionally identical. Figure 3a confirms that under the preferential vaccination scenario, the proportion of all individuals who are susceptible prior to the second season does not change even as the revaccination rate changes. This is as expected, because under the preferential vaccination scenario, the number of susceptible individuals is simply the total number of individuals in the network minus the number of individuals with natural immunity from the first season and minus the number of individuals vaccinated for the second season. Neither first season epidemic size nor vaccine coverage are functions of the revaccination rate.

On the other hand, we find that as the rate of revaccination increases, the proportion of susceptible individuals infected during the second season decreases regardless of whether individuals with natural immunity are vaccinated (Figure 3b). While the effect of revaccination is most pronounced in the universal vaccination scenario, a decrease in epidemic size is also evident under the preferential vaccination scenario. This result suggests that the increasing number of protected individuals does not fully explain the relationship between revaccination rate and epidemic size, and that higher revaccination rates also confer greater indirect protection to susceptible individuals. We present these two models of vaccination only to illustrate that revaccination has an impact on epidemic sizes even when the number of susceptible individuals is constant. Elsewhere in this study we focus only on the universal vaccination approach.

*Efficiency of Vaccination Schemes: Connectivity of Susceptible Individuals*

It is well-understood that public health measures such as vaccination can demonstrate an impact beyond those directly protected, to indirectly protect a larger community. The extent of this indirect protection, or *herd effect*, can be quantified in our model by comparing second-season epidemic sizes at elevated rates of revaccination to second-season epidemic sizes when the level of revaccination is equal to the overall coverage level (Figure 4), which is the level of revaccination expected if vaccination was not affected by previous vaccination status. Figure 4 shows that as revaccination increases, the population level efficacy of indirect protection increases as well.

To better understand why higher rates of revaccination indirectly protect susceptible individuals, we explore the degree (number of contacts or edges) of susceptible and protected (vaccinated or naturally immunized) individuals in the network. Vaccination in the first season is random, so the expected value of the average degree of nodes vaccinated in the first season is equal to the mean degree of the network (Figure 5a). However, infection during the first season and, therefore, natural immunity prior to the second season are not random with respect to degree, because higher degree nodes are more likely to be infected during the first season (Figure 5a) [32, 30]. Consequently, remaining susceptible nodes have disproportionately lower degree, on average, than both naturally immune individuals and individuals vaccinated during the first season (Figure 5a). In a population connected as an exponential random network, with complete revaccination, susceptible individuals in the second season have an average degree that is 15% smaller than that of susceptible individuals in the first season. However, when individuals vaccinated for the first season are not revaccinated, these nodes of average degree are, in effect, added to the set of nodes susceptible in the second season, thereby increasing the average degree of susceptible individuals (Figure 5b). This, in turn, decreases the strength of the herd effect in the network, because when high degree nodes are not protected either by vaccination or natural immunity, the epidemic is able to spread further [33-34].

*Robustness in Realistic Populations*

While we are not aware of any studies that focus on estimating influenza revaccination rates in large populations, we have been able to infer approximate revaccination rates from a variety of studies. Rates likely vary between populations, perhaps depending on factors such as access to vaccines and the overall rate of vaccination in the population. For example, data from a study of Medicare beneficiaries [35] indicates a revaccination rate of 93.4% between the 1998-99 and 1999-2000 influenza seasons with a 70% vaccine coverage rate; while Uddin et al. [36] surveyed college students and found a revaccination rate between the 2006-07 and 2007-08 seasons of 58.5% with an average coverage rate of 15.65%. Lastly, based on data reported in a large study of people 65 years and older in the Netherlands, we estimated rates of revaccination generally between 80 and 90% [17]. Adjusting these revaccination rates for varying coverage levels, we estimate plausible excess revaccination rates of 39% to 78% (Figure 6).

To explore the robustness of our findings, we test our hypotheses on an empirical contact network model as well as with realistic values of epidemiological parameters. The empirical network represents an urban population based on data for the city of Vancouver, British Columbia, and is built from age-specific, activity-based interaction patterns relevant to the spread of an influenza-like illness [37, 30]. Bansal et al. [30] find that while the Vancouver-based model has a higher density of contacts, age-specific contact patterns are captured well in the model compared with empirical data from studies on contact structure.

Based on this contact network, we assume age-specific vaccine efficacy and coverage rates based on the 2006 and 2011 influenza seasons in the United States, as well as levels of natural immunity based on empirical estimates from recent studies (details in Methods). Our findings in this scenario, based on the Vancouver host population and empirical parameters, are qualitatively similar to those found previously: increased revaccination decreases the proportion of individuals infected (Figure 6). While the impact of revaccination is more muted due to the use of an imperfect vaccine (as available in the two U.S. influenza seasons considered) with moderate levels of coverage, these results illustrate that revaccination does indeed reduce the burden on public health (for realistic estimates of $r'$), and has the capacity for a larger impact.

In addition, we conduct sensitivity analyses to assess whether our findings are robust with respect to partial natural immunity (Figure S1), vaccine efficacy (Figure S2) and vaccine coverage rates (Figures S3 and S4). We also studied whether assortative vaccination, a phenomenon that has been observed empirically [38, 39], affects the relationship between revaccination and epidemic size (Figure S5). In terms of network variables, we consider the impact of variation in network size (i.e. the size of the population) (Figure S6) and variance in node degree (Figure S7). In general, we find that the decrease in epidemic size due to revaccination is strongest when both vaccine efficacy and natural immunity are complete and when networks have degree distributions with high variance. In addition, lower vaccine coverage appears to increase the effect of revaccination on the total number of cases, highlighting that higher revaccination rates can be used to compensate for low coverage rates (e.g. the epidemic size for a vaccine coverage rate of 50% with no revaccination is equivalent to the epidemic size for a vaccine coverage rate of 30% with full revaccination) (Figure S4). Finally, we observed that the relationship between revaccination rate and epidemic size is robust to both assortative vaccination and network size. (All figures mentioned are in the Supplementary Materials).

**Discussion**

Using a mathematical modeling framework that accounts for the consequences of past epidemics on future disease outbreaks (Figure 1), we have considered the epidemiological impact of influenza revaccination. Our work suggests that implementing greater rates of revaccination may contribute to reduced outbreak sizes (Figure 2), both by reducing the overall number of individuals who are susceptible by using vaccine more efficiently (Figure 3), and by increasing the extent to which more connected individuals are protected (Figures 4 and 5). We also show that similar results are obtained in populations with more realistic contact structure, with empirical estimates of natural immunity levels and less than ideal vaccine coverage and efficacy levels from recent influenza outbreaks (Figure 6). While we have focused our attention on the impact of revaccination on the total incidence of influenza, we expect similarly positive results for other metrics of public health impact (e.g. peak incidence and outbreak duration).

The process described in this study can be thought of as a partial fragmentation of the contact network by first season vaccination and, especially, the random vaccination of highly-connected individuals. Infection then spreads, working its way through the most connected parts of the network, but its path is constrained by first season vaccination. When previously vaccinated individuals are not revaccinated (i.e. when *r* is low), previously protected fragments of the network are made vulnerable. This compromises the strength of herd protection, thereby creating new paths through the population along which infection during the second season can spread, leading to larger second season epidemic sizes.

More generally, we have demonstrated that mathematical models to develop and test influenza vaccination schemes should take into account prior vaccination status, as the distribution of vaccination in a population, even if it is random, can shape patterns of natural immunity. In turn, patterns of natural immunity are not random and drive the frailty, which is the extent to which highly connected individuals are at risk of infection, of the host population [32, 30]. These findings also reinforce previous work that highlights the need for shifting influenza control strategy with the epidemiological structure of a population, and targeting those most likely to be infected [30]. While this study focuses on human influenza in particular, the population-level consequences of revaccination rates may be relevant to research on other infectious disease systems with complex multi-strain natural and vaccine-induced immunity dynamics, such as dengue and even swine influenza or foot and mouth disease in livestock [40-42]. This study is relevant to public and animal health policy because it contributes a population-level perspective to debates about the merits of repeated influenza vaccination.

One of the limitations of this study, however, is that in real populations, there is no corresponding first season of seasonal influenza in which natural immunity does not exist, as is assumed here. A better understanding of the longitudinal distribution and properties of natural and vaccine-induced immunity to influenza in empirical contact networks would significantly enhance modeling efforts. Also, this model assumes that the host population is closed and that the network structure is constant across both seasons. Presumably those assumptions become less tenable as models take into account more seasons in series. However, the mechanism of indirect protection we have identified relates to the average connectivity of immune and susceptible individuals, not necessarily their particular place in the network. Therefore, future research should address not only how contact networks change over time, but also the extent to which the degree of individuals remains consistent over time. Finally, this model does not explicitly take into account viral evolution (viral evolution is an implicit factor only to the extent that vaccine failure is due to antigenic drift). Presumably, the validity of the assumption that vaccine-induced immunity is season-specific and that natural immunity is effective across seasons varies depending on the particular pair of seasons under consideration.

While vaccinating individuals with large numbers of contacts is arguably advantageous regardless of the individual's immunological history, this study indicates that it may be especially important to vaccinate individuals who, by virtue of their occupation or living arrangements, are likely to have a high number of contacts and who have been vaccinated previously. Due to past vaccination, such individuals may be less likely to have natural immunity and, if not vaccinated, could infect large numbers of contacts. Some health care systems send reminders to people vaccinated in previous years to be vaccinated for the upcoming influenza season [43, 35]. Practically, this is an effective practice because past vaccination is a strong predictor of future willingness to be vaccinated, but this study shows that this health care intervention may have population-level benefits beyond that of simply increasing vaccine coverage. The results of our study demonstrate that policy debates about repeated influenza vaccination and the related topic of universal vaccination should take into account disease ecology and, especially, herd immunity considerations, not just immunological and public health implementation considerations.

**Methods**

*Defining Revaccination*

We define the rate of revaccination (*r*) as the proportion of vaccine recipients in one influenza season who are also vaccinated in the following season. The revaccination rate is easiest to conceptualize if populations are closed and the level of vaccine coverage is constant across both seasons. As vaccine coverage approaches 100%, the revaccination rate also converges to 100%. However, if the overall vaccination coverage is less than or equal to 50%, the range of theoretically possible revaccination rates is 0% to 100%. If vaccination were random with respect to previous vaccination status, the expected revaccination rate would be equal to the vaccination coverage. To account for this, we additionally define $r'$ to be the excess revaccination rate, which measures revaccination beyond what is expected at random,

$$r' = \frac{r - C}{1 - C}$$

where $r$ is the absolute rate of revaccination and $C$ is the vaccine coverage.

*Population Model*

We simulated epidemics on computationally-generated contact network structures. Computationally-generated theoretical networks allow us to systematically study the epidemiological consequences of network structure. The network structure used in this study is an exponential random network, with the number of contacts (i.e. degree) per individual sampled from a geometric distribution, and connected randomly. We assume an average degree of 10 contacts per individual and a network size of 5,000 nodes. (The impact of these choices is studied in the sensitivity analyses.) While much remains unknown about contact networks in real populations, Bansal et al. [44] found that contact networks derived from empirical data correspond more closely to exponential random network structures than other common network types.

*Epidemiological Two-Season Model & Vaccination*

We model first-season vaccination with single doses of influenza vaccine by removing select individuals and all their connections from the network. Individuals to be vaccinated are chosen randomly, and the size of the population to be vaccinated (and thus fully protected against influenza) is $CE$, where $C$ is the vaccine coverage rate, and $E$ is the vaccine efficacy. We call the set of individuals who are vaccinated in the first season, $V_1$, and note that these individuals are only protected for the first season (as influenza vaccine-induced immunity is temporary).

To model the first season outbreak, we perform Monte Carlo simulations for a susceptible–infected–recovered (SIR) epidemic model with a single initial infected case and per-contact transmissibility, $T_1$, on all susceptible individuals in the networks specified above. Once infected, a node cannot be reinfected during the same season, and unlike with vaccination, will have resistance to infection during the subsequent season (natural immunity). This is a reasonable assumption because natural immunity is thought to induce a stronger, longer lasting immune response than vaccines, and provide better cross-protection across strains [45-48].

Second-season vaccination is modeled similarly, except the identity of vaccinated individuals is no longer completely random. Based on the revaccination rate, $r$, a proportion $r$ of the vaccinated group ($V_1$) is vaccinated first. The remaining vaccine supply ($C-rV_1$) is distributed randomly among the rest of the population.

The second season outbreak is also modeled with a Monte Carlo SIR model, and an independent transmissibility $T_2$. Infection in the second season is allowed in all susceptible individuals (that is, those individuals who do not have natural immunity from the first season and those who have not been vaccinated immediately prior to the second season). Second season outbreaks are only considered in cases when a large epidemic occurs in the first season. The model assumes constant demography and constant network structure over the course of the two seasons. Public health burden is measured in terms of the proportion of the population infected in the case that there is a large epidemic in the second season ("second-season size of epidemic").

*Realistic Parameters*

In the robustness analysis on the Vancouver urban network [37, 30], we divide the population into four age classes: Ages 0-4, 5-18, 19-64, and ≥65. The vaccine efficacies for each of these age groups were 60%, 60%, 70%, and 50%, respectively, and were primarily based on clinical trials of influenza vaccines and meta-analyses thereof [49-55, 7]. We also implement two vaccine coverage scenarios. For a scenario based on the 2011-2012 influenza season in the United States, the age-specific vaccine coverage levels are 55%, 45%, 40%, and 70%, respectively [56]; for a scenario based on coverage levels from 2006 in the United States, age-specific vaccine coverage levels are 33%, 16%, 21%, and 65%, respectively [57-58]. Revaccination rates are applied to each age class such that the excess revaccination is equal across age classes (i.e. differences in vaccine coverage are taken into account.) While the relationship between immune response and future protection is not well quantified, the efficacy of natural immunity, $Q$, for all age groups is assumed to be 80% [59-60, 45, 48]. In this case, natural immunity is implemented in a manner similar to vaccination, so that 80% of those infected in the first season are assumed to be fully protected against infection, while the remaining 20% are not protected at all.

The methods for the remaining sensitivity analyses are described in the Supplemental Materials.

**Acknowledgements**

The authors thank Lauren Ancel Meyers for the Vancouver contact network dataset and Sarah Kramer for helpful discussions on vaccine efficacy. This work was supported by the Howard Hughes Medical Institute Precollege and Undergraduate Science Education Program; and the RAPIDD Program of the Science & Technology Directorate, Department of Homeland Security and the Fogarty International Center, National Institutes of Health.

**Figures**

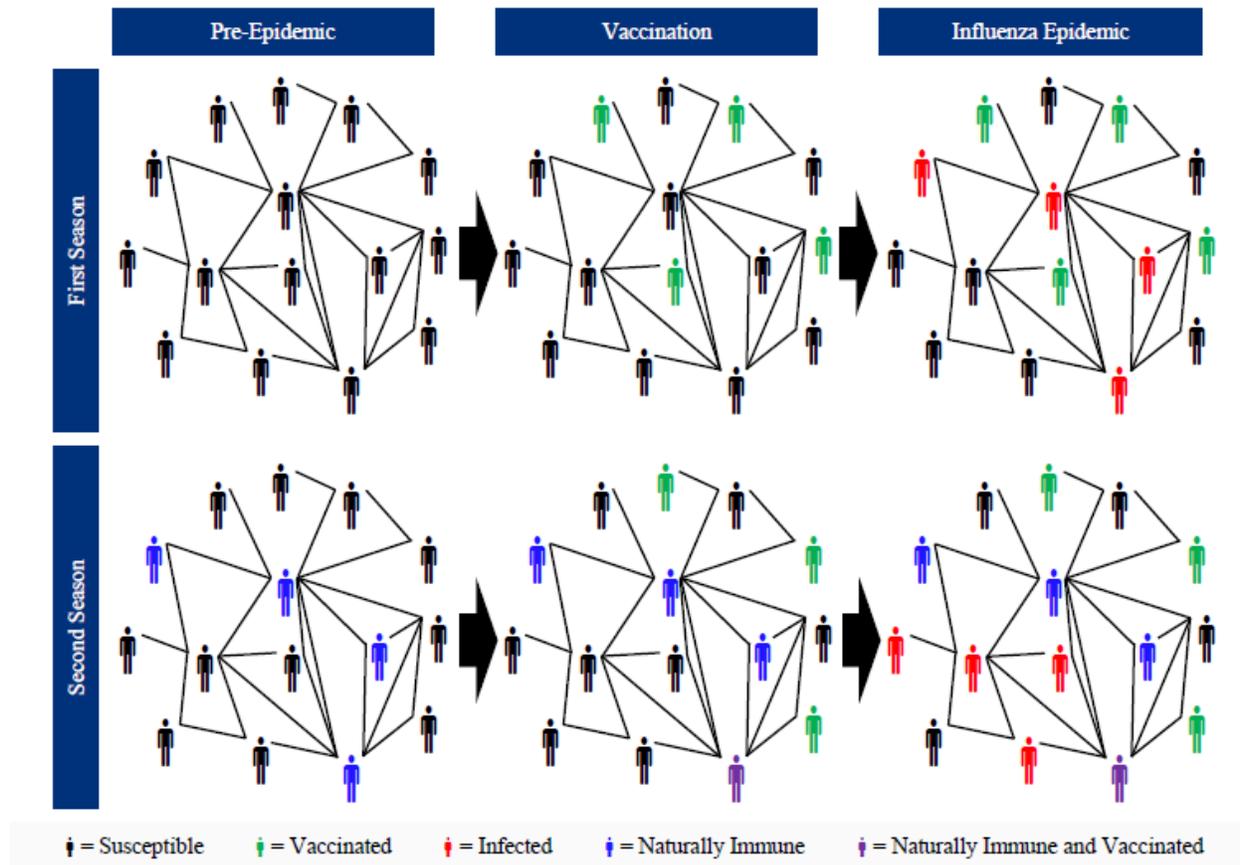

Figure 1. Schematic representation of a two-season contact network model for seasonal influenza. Individuals make up nodes in the contact network, and contacts between individuals are represented by network edges. This heuristic representation assumes that both natural immunity and vaccine efficacy are complete ($E=1.0$ and $Q=1.0$). The scenario in which there is no revaccination ($r=0$) is illustrated here. The universal revaccination scheme in which naturally immune individuals may be randomly selected for second season vaccination is used (note the individual, in purple, who is both naturally immune and vaccinated).

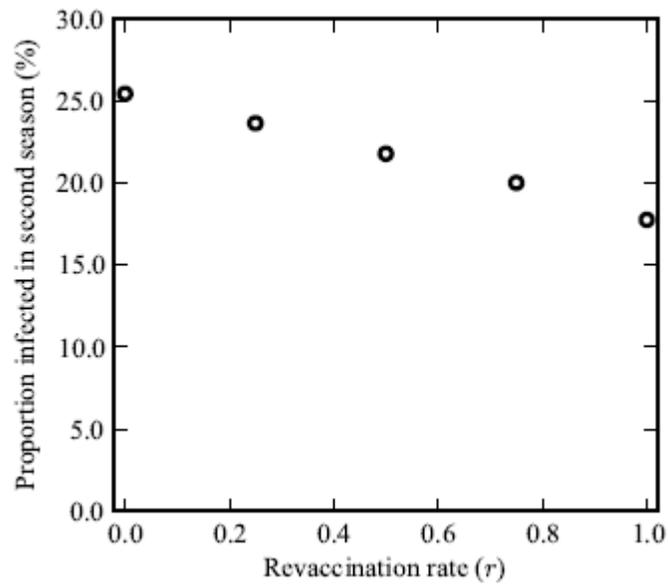

Figure 2. Second season epidemic size decreases as revaccination increases. This figure is based on results from 2000 simulated second season large epidemics on a single 5000 node exponential random network with $T_1=.09$, $T_2=.18$, $E=1.0$, $Q=1.0$, and $C=.25$.

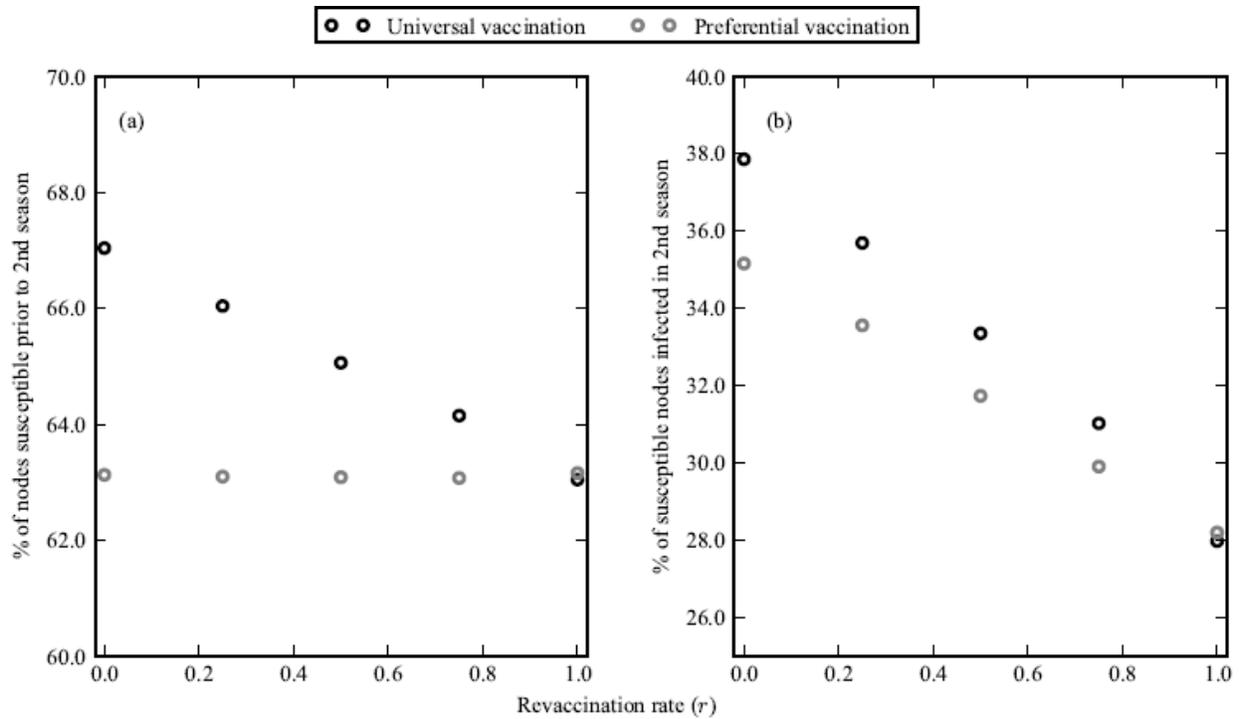

Figure 3. a) The proportion of all individuals susceptible prior to the second season decreases as the revaccination rate increases when the universal vaccination scenario is used (black), but is held constant under the preferential vaccination scenario (gray). b) The proportion of susceptible individuals infected during the second seasons decreases as the revaccination rate increases regardless of whether the universal vaccination scenario (black) or the preferential vaccination scenario is used (gray). This figure is based on results from 2000 simulated second season large epidemics on a single 5000 node exponential random network with $T_1=.09$, $T_2=.18$, $E=1.0$, $Q=1.0$, and $C=.25$. This is the only figure that contrasts results from the universal and preferential vaccination schemes. Elsewhere, only universal vaccination is shown.

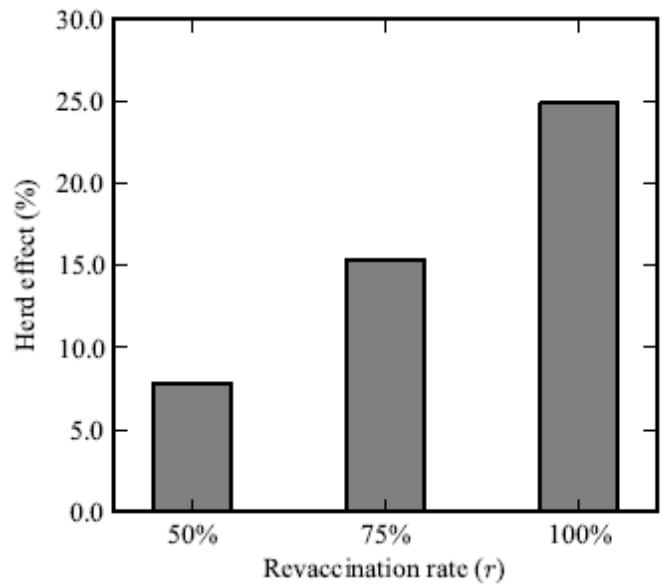

Figure 4. This figure shows the strength of the herd effect (or, indirect protection) at different levels of revaccination. Efficacy is calculated as 1-*RR*, where *RR* is the relative risk, calculated as the ratio of incidence among unvaccinated individuals at the specified rate of revaccination and incidence among unvaccinated individuals when the revaccination rate is equal to the vaccine coverage (i.e. *r'=0*). This figure is based on results from 2000 simulated second season large epidemics on a single 5000 node exponential random network with $T_1$=.09, $T_2$=.18, *E*=1.0, *Q*=1.0, and *C*=.25.

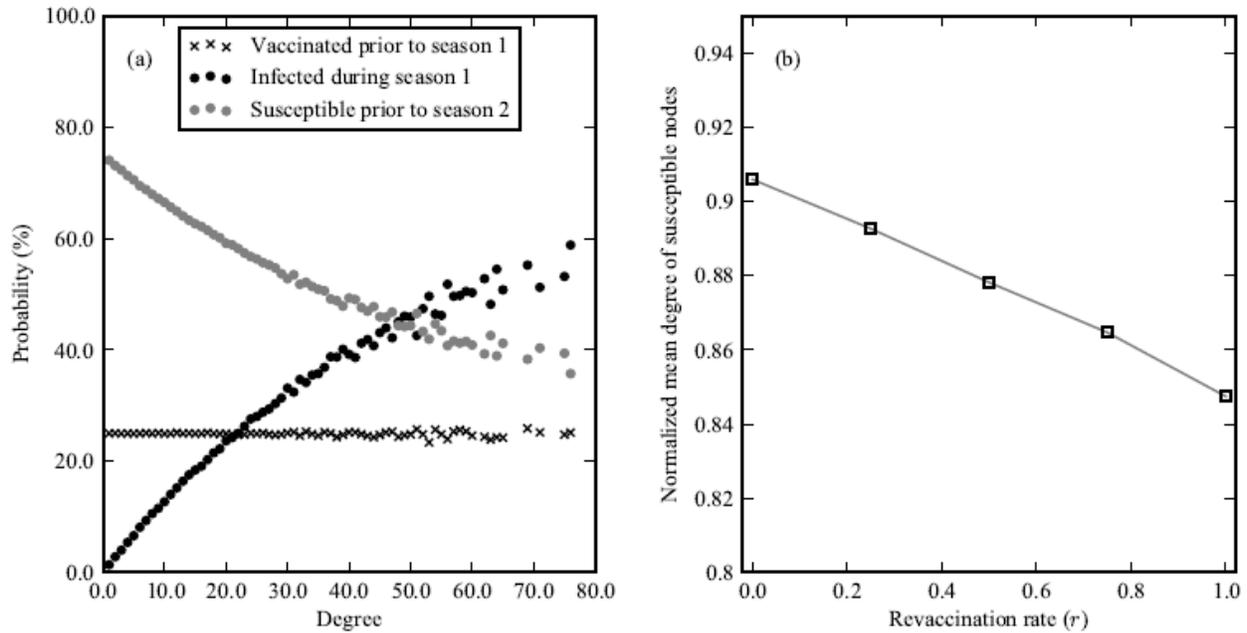

Figure 5. a) Estimated probability of first season vaccination is constant regardless of degree (black x). Nodes of higher degree are more likely to be infected during the first season epidemic and therefore more likely to have natural immunity (black circles). Consequently, lower degree nodes are more likely to be susceptible prior to the second season epidemic (gray circles). The probabilities shown here are calculated for the 0% revaccination level, but the qualitative patterns are similar across all revaccination rates. We note that this panel does not reflect the degree distribution of the network. b) At higher revaccination rates, the mean degree of second season susceptible nodes decreases. Here, the mean degree of susceptible nodes is normalized by the network's mean degree. This figure is based on results from 2000 simulated second season large epidemics on a single 5000 node exponential random network with $T_1=.09$, $T_2=.18$, $E=1.0$, $Q=1.0$, and $C=.25$.

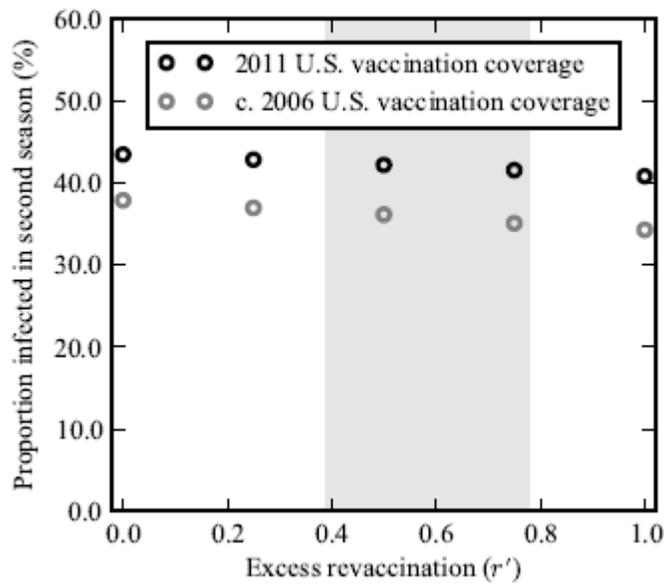

Figure 6. Second season epidemic size decreases as revaccination increases. This figure shows results from 2000 simulated second season large epidemics on an age-structured network representative of contact patterns in Vancouver, British Columbia, with partial natural immunity ($Q$=80%) and age-specific vaccine efficacies and coverage rates (see Methods). Two vaccination scenarios are illustrated here, one based on vaccine coverage levels in the United States during the 2011-2012 influenza season and the other based on vaccine coverage levels circa 2006 in the United States. The light gray box corresponds to values of excess revaccination calculated from empirical studies of vaccination (see Results).

**Increasing Herd Immunity with Influenza Revaccination**

Eric Mooring and Shweta Bansal

**Supplemental Materials**

*Sensitivity Analysis—Model Variables*

We conducted numerous simulations to test how changing key parameters in the model affected the relationship between revaccination rate and epidemic size. Unless otherwise noted, 1000 iterations were conducted for each combination of values. We found that when natural immunity is stronger, increasing the rate of revaccination has a larger effect on the size of second season epidemics (Figure S1). This finding makes intuitive sense because in the limit where there is no natural immunity across seasons, each season can be thought of as independent and equivalent. As expected, we also found smaller second seasons when natural immunity is stronger, regardless of level of revaccination.

Additionally, we considered the consequences of changing efficacy of vaccination. The effect of revaccination on epidemic size is strongest when vaccine is fully protective (Figure S2). Regardless of the degree of revaccination, we found higher second season epidemic sizes when vaccine efficacy was greater, because greater vaccine efficacy in the first season limited the degree of natural immunity in the second season and *vice versa*. For this reason, we plotted the summed first and second season epidemic sizes.

Next, we considered the effect of changing the level of vaccine coverage. This sensitivity analysis is complicated by the fact that the level of vaccine coverage is not wholly independent of the rate of revaccination. Therefore, we analyzed the effect of changing rates of excess revaccination, which is measured as

$$r' = \frac{r - C}{1 - C}$$

where $r$ is the absolute degree of revaccination and $C$ is the vaccine coverage. This can be thought of as the extent of revaccination beyond that which would be expected by chance ($C$) normalized by the magnitude of the range of possible excess revaccination (1-$C$). Figure S3 indicates that the effect of revaccination is greater at lower vaccine coverage levels, but this effect is largely because at lower vaccine coverage levels, the range of possible excess revaccination rates is larger. Specifically, figure S4 plots the same data but plots the absolute revaccination ($r$) as opposed to the excess revaccination ($r'$) on the x-axis. The effect of increasing revaccination at high vaccine coverage levels now appears magnified. Because the degree of vaccine coverage affects the first season epidemic size and hence the degree of natural immunity in the population, as with the sensitivity analysis on vaccine efficacy, we plot summed first and second season epidemic sizes.

Finally, we investigated whether assortativity with respect to vaccination affects the relationship between revaccination and epidemic size. There does not appear to be any systematic effect on the relationship, but assortativity affects the size of the first season epidemic and thus the amount of natural immunity in the community, so we therefore present summed first and second season epidemic sizes (Figure S5).

Assortative vaccination assignments were generated based on the algorithm of Salathé and Khandelwal [1]. Assortative first season vaccination assignments were created by first randomly assigning vaccination at a given coverage level and calculating the assortativity coefficient $\rho$. Next, two nodes with opposite vaccination status were selected and their assigned vaccination statuses were swapped; $\rho$ was recalculated and if $\rho$ was increased by the swap then it was preserved. If not, the pair of nodes reverted to their original assigned vaccination status. This process was repeated until the target assortativity was reached. In this sensitivity analysis, target assortativities of 0.05, 0.10, 0.15, 0.20, and 0.25 were used. The values span values observed in an empirical study of seasonal influenza vaccination on a contact network [2]. To generate second season vaccination assignments, the first season assignments and a given level of revaccination were used to generate preliminary second season assignments. Then, as before, the assortativity was boosted as needed by randomly selecting pairs of nodes with opposite second season vaccination status, swapping statuses, and recalculating $\rho$. But now, all pairs of nodes had to share the same first season vaccination status. (This process was unnecessary when revaccination was complete, because the first season assignments were reused and already met the target level of assortativity.) We did not investigate having the target level of assortativity vary between seasons.

*Sensitivity Analysis—Network Variables*

We conducted simulations to determine the effect of changing the structure of the network on the relationship between the degree of revaccination and second season epidemic size. To test the effect of network size, we computationally generated exponential random networks with 250, 500, 5,000, 15,000, and 25,000 nodes. While the degree distributions were drawn from theoretical distributions with mean degree of 10, the networks actually differ slightly in mean degree, which influenced the size of first seasons and, in turn, the extent of natural immunity. Therefore, for the same reason as in the assortativity sensitivity analysis, the response variable shown in this analysis is the summed first and second season epidemic size. Figure S6 shows the results of simulations on ten networks of each size. Note that simulations using the smaller networks exhibit greater variance in epidemic size. However, there does not appear to be a systematic effect of network size on the relationship between revaccination rate and epidemic size.

Finally, we assessed the effect of changing variance of the degree distribution on the relationship between revaccination rate and epidemic size. We computationally generated negative binomial networks from negative binomial distributions with mean degree 10 and variances 10, 13.5, 18, 22.5, 36, and 90. The negative binomial distributions were shifted rightward by one unit so that zero was excluded from the degree distribution. We found that the effect of increasing revaccination was strongest when variance in degree was greatest (Figure S7). This finding confirms our explanation that non-randomness of natural immunity with respect to degree partially underlies the effect of revaccination on epidemic size. Because the variance of the degree distribution affects first season epidemic sizes, summed first and second season epidemic sizes are plotted in figure S7.

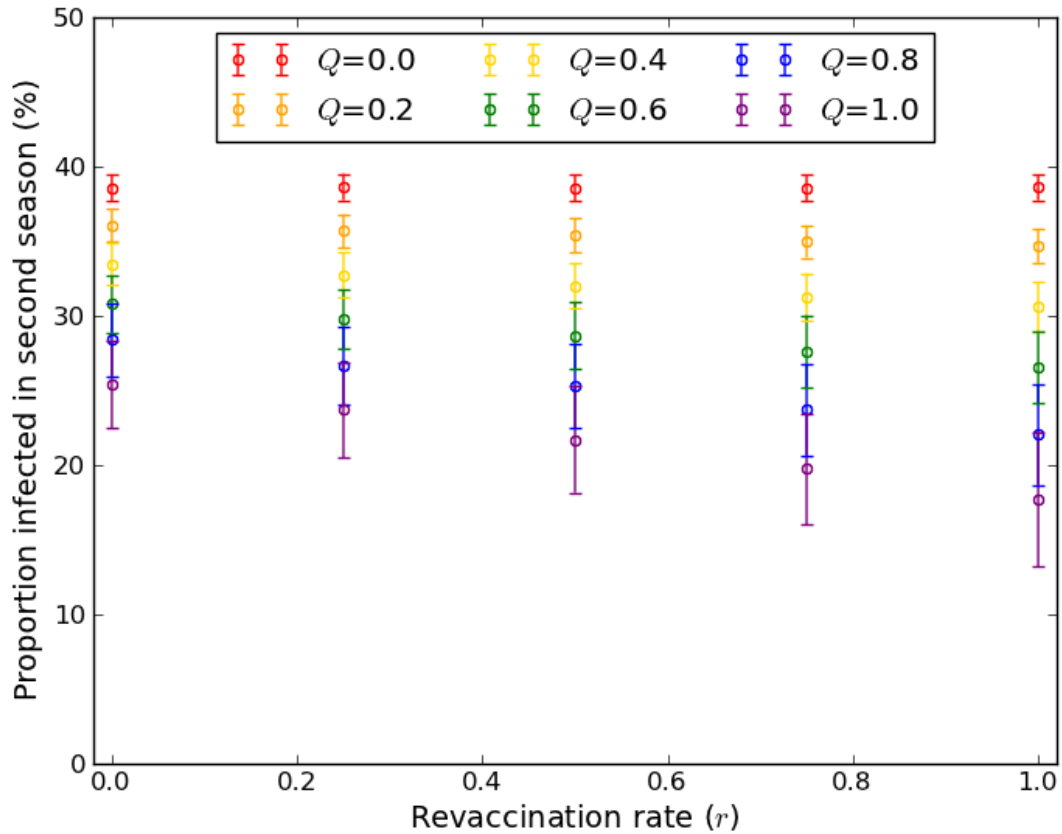

Figure S1. When natural immunity is stronger, the effect of revaccination on mean second season epidemic size is more pronounced. This figure is based on results from 1000 simulated second season large epidemics on a single 5000 node exponential random network with $T_1=.09$, $T_2=.18$, $E=1.0$, and $C=.25$. Error bars are ± 1 standard deviation.

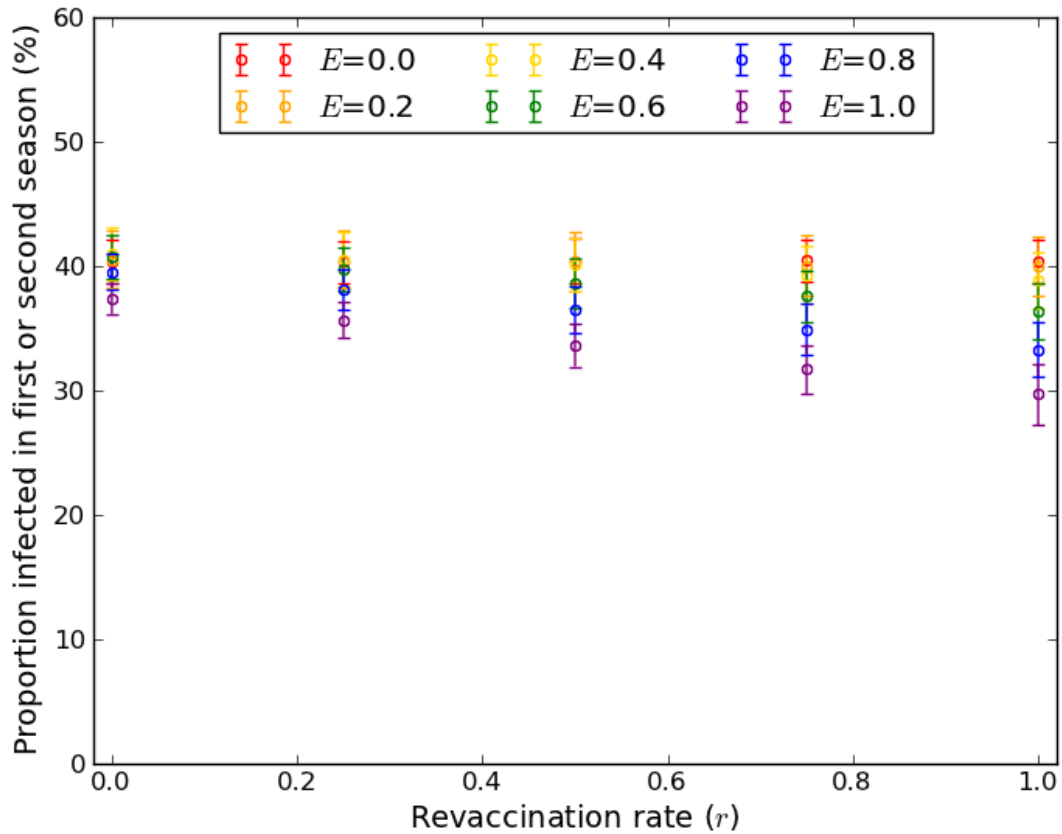

Figure S2. When the efficacy of vaccination is greater, the effect of revaccination on epidemic size is more pronounced. This figure is based on results from 1000 simulated second season large epidemics on a single 5000 node exponential random network with $T_1=.09$, $T_2=.18$, $Q=1.0$, and $C=.25$. Error bars are ± 1 standard deviation.

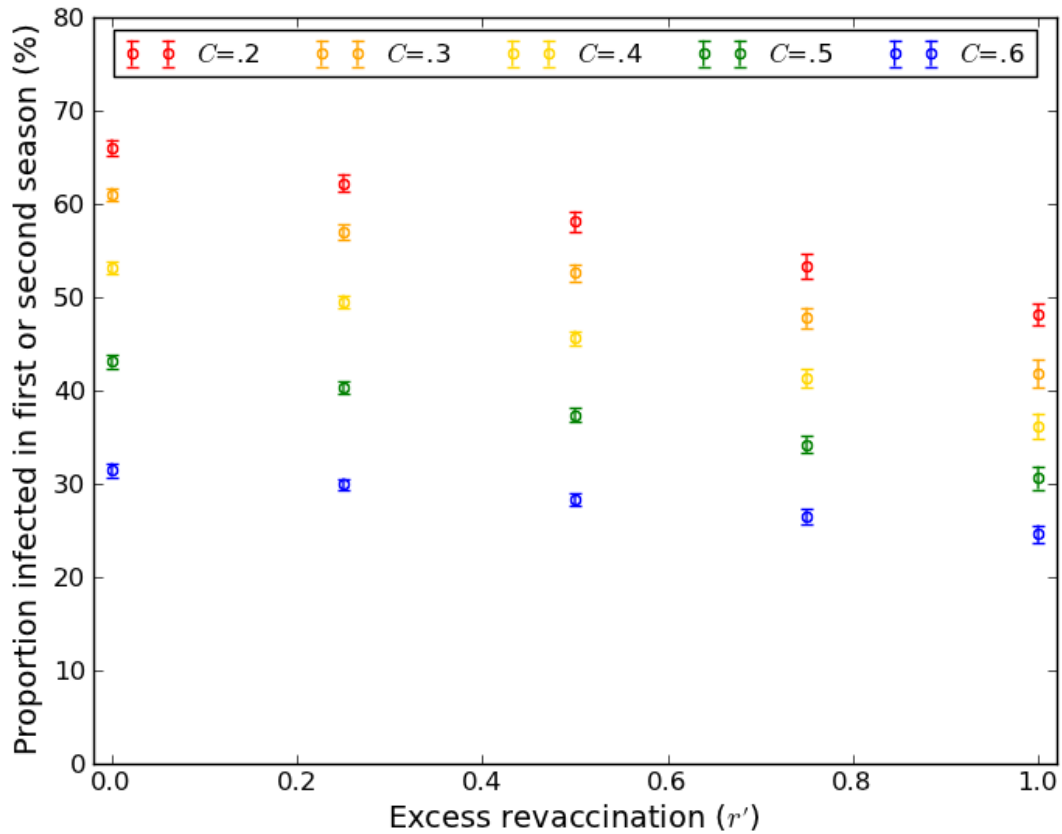

Figure S3. When plotted as a function of excess revaccination, lower vaccine coverage appears to increase the effect of revaccination on epidemic size, at least for the range of vaccine coverage levels simulated here. (However, see figure S4.) This figure is based on results from 1000 simulated second season large epidemics on a single 5000 node exponential random network with $T_1$=.17, $T_2$=.6, $E$=1.0, and $Q$=1.0. Error bars are ± 1 standard deviation.

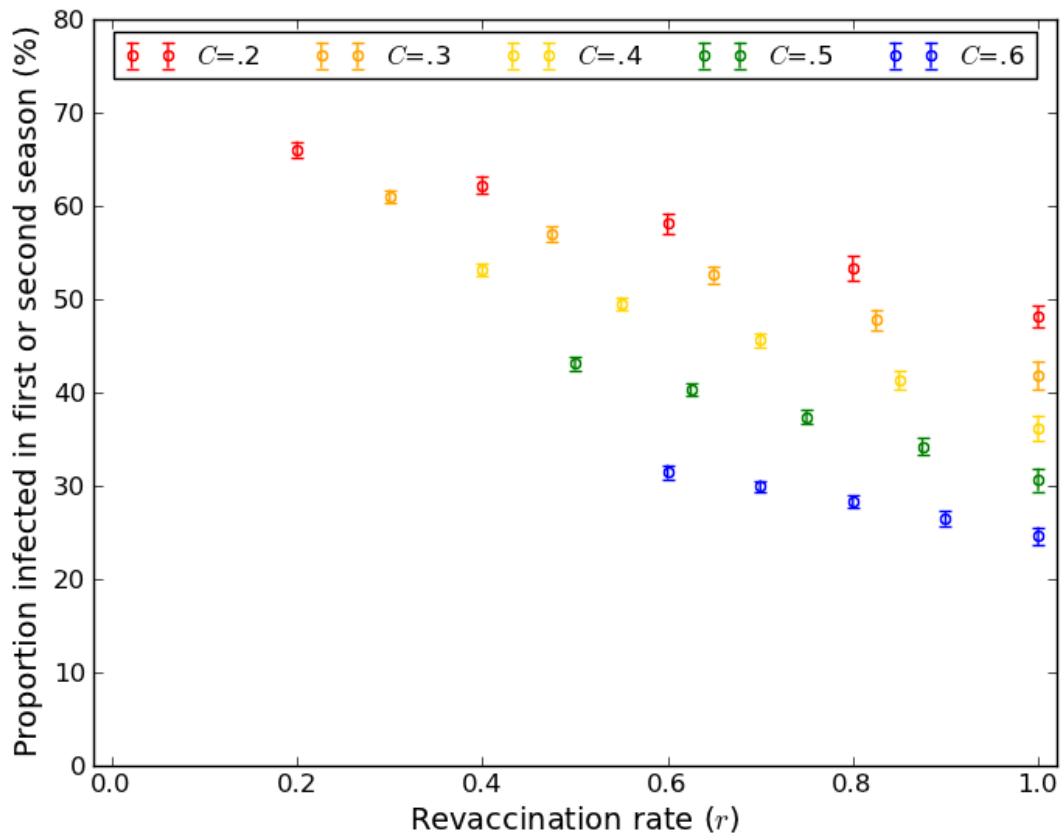

Figure S4. This figure plots the same data as figure S3, but as a function of the absolute degree of revaccination. Note that the series of points for each vaccine coverage level appear more parallel in this figure than in figure S3. This figure is based on results from 1000 simulated second season large epidemics on a single 5000 node exponential random network with $T_1=.17$, $T_2=.6$, $E=1.0$, and $Q=1.0$. Error bars are ± 1 standard deviation.

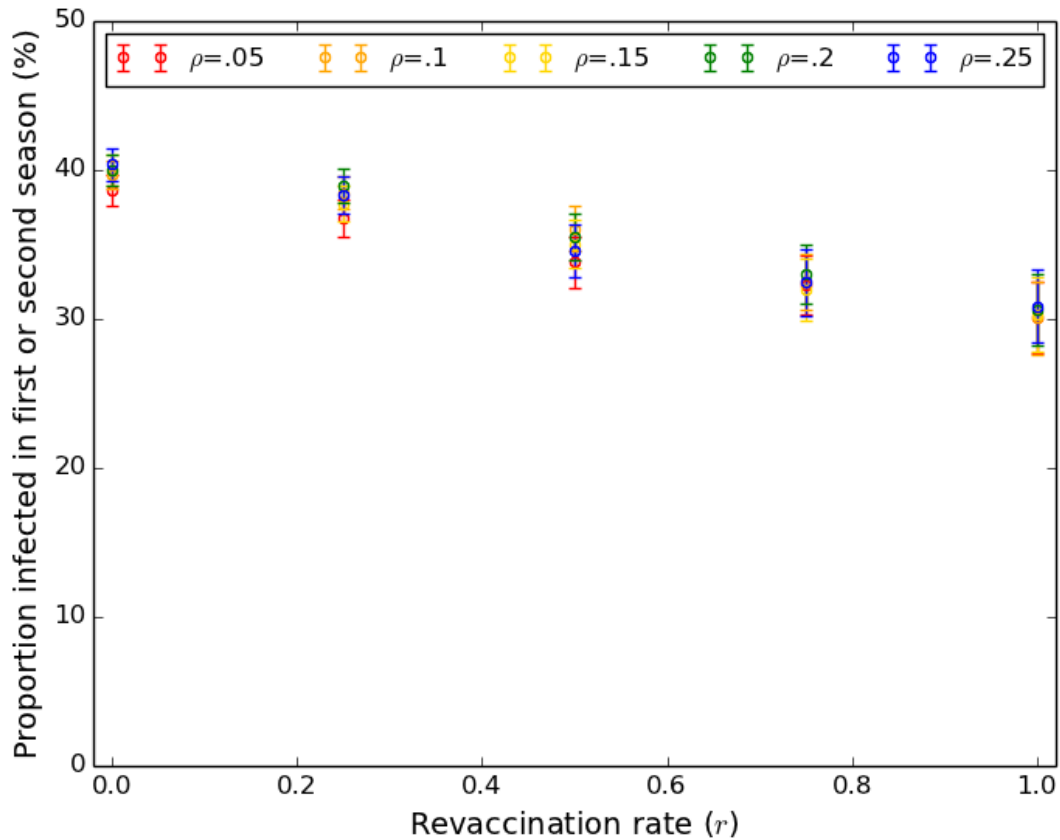

Figure S5. It does not appear that vaccine assortativity, $\rho$, modifies the relationship between revaccination and epidemic size. Each point is based on 2000 simulated first and second season epidemics, but unlike in the other model simulations in this article, the same vaccine assignments were used for all simulations at each combination of assortativity and revaccination. Epidemics were simulated on a single 5000 node exponential random network with $T_1=.09$, $T_2=.18$, $E=1.0$, $Q=1.0$, and $C=.25$. Error bars are ± 1 standard deviation.

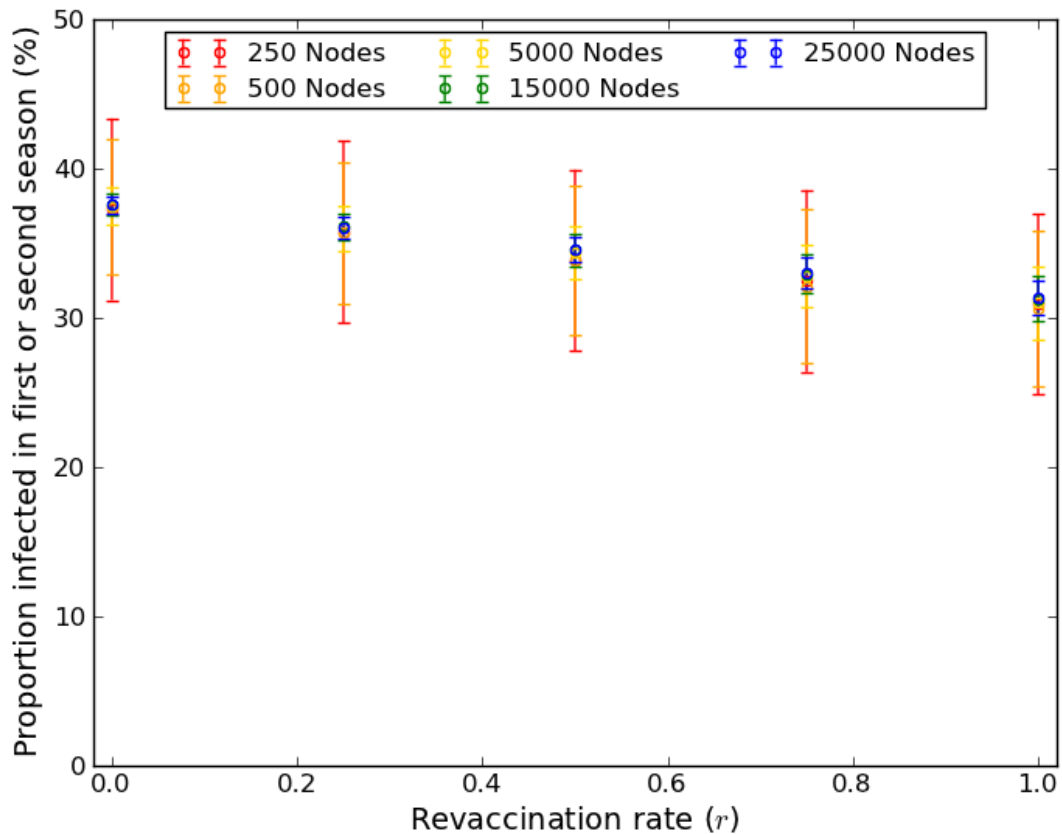

Figure S6. There does not appear to be a major effect of network size on the relationship between revaccination and epidemic size. Simulations on ten different networks of each listed size were conducted. This figure is based on pooling results from 500 simulations on each exponential random network with $T_1$=.09, $T_2$=.18, $E$=1.0, $Q$=1.0, and $C$=.25. Error bars are ± 1 standard deviation.

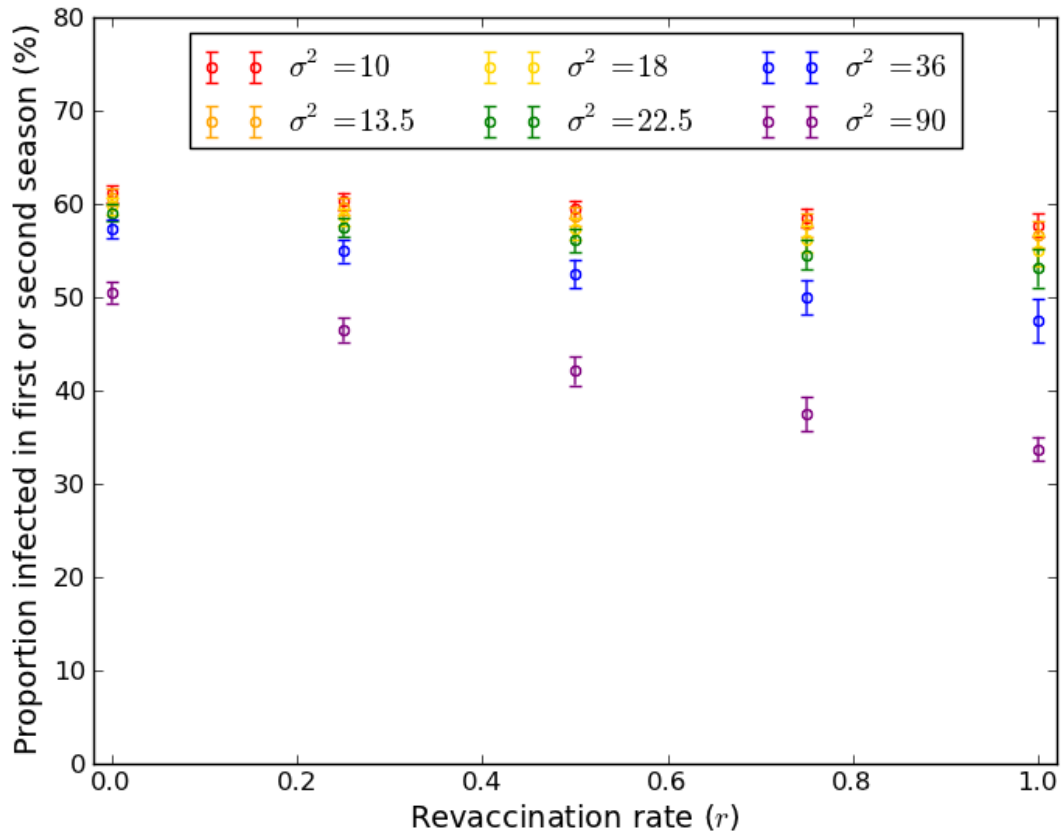

Figure S7. When the variance of the degree distribution is greater, the effect of revaccination on epidemic size is more pronounced. This figure is based on results from 1000 simulated second season large epidemics on each 5000 node, mean degree 10 negative binomial random network with $T_1=.13$, $T_2=.17$, $E=1.0$, $Q=1.0$, and $C=.25$. Error bars are ± 1 standard deviation.